\documentclass[twocolumn,showpacs,preprintnumbers,amsmath,amssymb,prb]{revtex4}

\usepackage{graphicx}
\usepackage{dcolumn}
\usepackage{bm}

\begin{document}


\title{Enhanced current quantization in high frequency electron pumps in a perpendicular magnetic field}

\author{S.~J.~Wright$^{1,3}$, M.~D.~Blumenthal$^{2,1}$, Godfrey Gumbs$^4$, A.~L.~Thorn$^{1}$, M.~Pepper$^1$, T.~J.~B.~M.~Janssen$^2$, S.~N.~Holmes$^{3}$, D.~Anderson$^{1}$, G.~A.~C.~Jones$^{1}$, C.~A.~Nicoll$^{1}$, D.~A.~Ritchie$^{1}$\\
}

\affiliation{
${}^1$\footnotesize Cavendish Laboratory, University of Cambridge, J. J. Thomson Avenue, Cambridge CB3 0HE, UK.
${}^2$\footnotesize National Physical Laboratory, Hampton Road, Teddington TW11 0LW, UK.
${}^3$\footnotesize Toshiba Research Europe Ltd, Cambridge Research Laboratory, 208 Science Park, Milton Road, Cambridge CB4 0WE, UK.
${}^4$\footnotesize Department of Physics and Astronomy, Hunter College of the City University of New York, 695 Park Avenue, New York, New York 10065 USA.
}

\date{\today}

\begin{abstract}

We present experimental results of high frequency quantized charge pumping through a quantum dot formed by the electric field arising from applied voltages in a GaAs/AlGaAs system in the presence of a perpendicular magnetic field $B$. Clear changes are observed in the quantized current plateaus as a function of applied magnetic field. We report on the robustness in the length of the quantized plateaus and improvements in the quantization as a result of the applied $B$ field.

\end{abstract}

\pacs{Valid PACS appear here}
\maketitle

Single electron turnstiles\cite{geerligs1990, andregg1990, kouwenhoven1PBI} and pumps\cite{pothier1PBI} were the first devices demonstrating electron transfer in a controlled manner. By periodically modulating the tunnel barriers electrons were moved through a quantum dot (QD). In these single electron tunneling (SET) devices, the transport of individual electrons can be controlled by means of an externally applied periodic signal at a certain frequency $f$. The generated current $I$ is determined by the integer number $n$ of electrons that are transported in each cycle and the frequency $f$ at which the pump is operated, such that $I_\mathrm{pump} = n e f$ (with $e$ the elementary charge). Keller \emph{et al}\cite{keller} operated a 7-junction electron pump with an error per pumped electron of 15 parts in 10$^{9}$. In such systems, quantized current is observed with frequencies limited to 20$\,$MHz which allows enough time for the stochastic tunnel process to take place. Much work has been carried out on surface acoustic waves (SAWs) on a piezoelectric GaAs substrate operating around 3$\,$GHz where quantized acoustoelectric current is observed through a single split gate configuration\cite{Shilton1996,Thornton1986}. More recently the realization of gigahertz charge pumping\cite{Blumenthal2007} has led to a marked increase in the total pumped current when compared with earlier pumps and turnstiles. The ease of operation\cite{kaestner:153301} and robustness\cite{kaestner:192106} of these devices allows for their potential application in metrology, quantum computation, single photon production and integrated single electron circuits.

In this paper we will study the effects of a perpendicular magnetic field on the ability to accurately capture and eject electrons in a single-electron pump. We find that the magnetic field leads to a significant improvement in the accuracy and robustness of current quantization. This is useful because single-electron pumps are a promising candidate for defining an electrical standard for current. An error of less than one part in $10^6$ with a current of at least one nanoampere is required. Application of a magnetic field is shown to aid in obtaining the required accuracy. Also, performing such studies may allow us to gain a better understanding of the pump operation which could lead to further improvements in quantization.

\begin{figure}[h]
\includegraphics[width=0.45\textwidth]{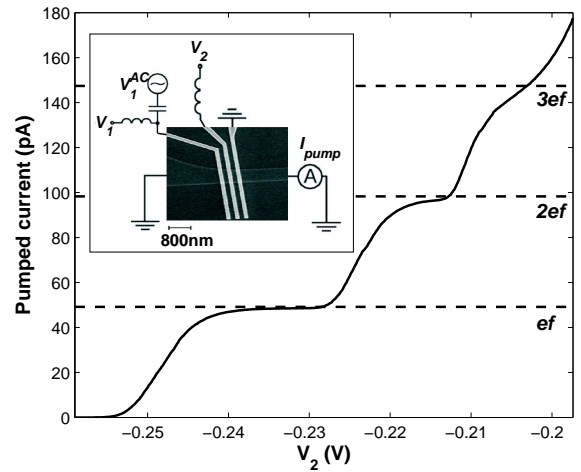}
\caption{\label{fig:plat} Pumped current as a function of $V_{2}$. Here, $V_1=-0.140V$ and $V_1^{AC}=306.7\,$MHz at $-12\,$dBm. Plateaus are visible at currents corresponding to a quantized number of electrons pumped per cycle. The expected values for these plateaus are shown as dashed lines in the figure. Inset is an SEM image of the device, including a schematic of the electrical connections.}
\end{figure}

In order to periodically form a QD our device uses two surface finger gates with applied voltages $V_{1}$ and $V_{2}$ to define a potential in a longitudinal direction and a shallow etch to impose a fixed transverse potential well \cite{Blumenthal2007,kaestner:153301}. An SEM image of the device can be seen in the inset of Fig.~\ref{fig:plat}. The narrow quantum wire of lithographic width $w=700 \,$nm was defined by shallow wet chemical etching on a Si doped AlGaAs/GaAs heterostructure with a two-dimensional electron gas (2DEG)  $90\,$nm below the surface. The wafer had a mobility of $120\,$ m$^2/$Vs and an electron density of $1.7 \times 10^{15}\,$m$^{-2}$ at $1.5\,$K. The wire was etched to a depth of $25\,$nm in an etchant of $\mathrm{HCl:H_{2}O_{2}:H_{2}O} = 1:4:100$. At each end of the wire are regions of 2DEG where Ohmic contacts are made.  Three Ti($10\,$nm)/Au($20\,$nm) finger gates with a width of $100\,$nm and a pitch of $250\,$nm were defined over the etched channel with electron beam lithography and metal gate evaporation.

A static QD is induced by applying negative DC voltages $V_1$ and $V_2$, setting the average barrier height of the QD. To induce the action of pumping, a sinusoidal signal from an RF generator $V_{1}^{AC}$ is added to the DC voltage $V_{1}$ using a bias tee, shown schematically in the inset of Fig.~\ref{fig:plat}. The frequency of the applied AC signal was set to $f=306.7\,$MHz with a power of $-12\,$dBm. From a study of the pinchoff characteristics for $V_{1}$ as a function of applied RF power we determined the power to voltage amplitude conversion, giving $\sim$80$\,$mV at $-12\,$dBm. All measurements presented in this work were performed in a dilution refrigerator at a temperature of $\sim$50$\,$mK.

\begin{figure}[h]
\includegraphics[width=0.48\textwidth]{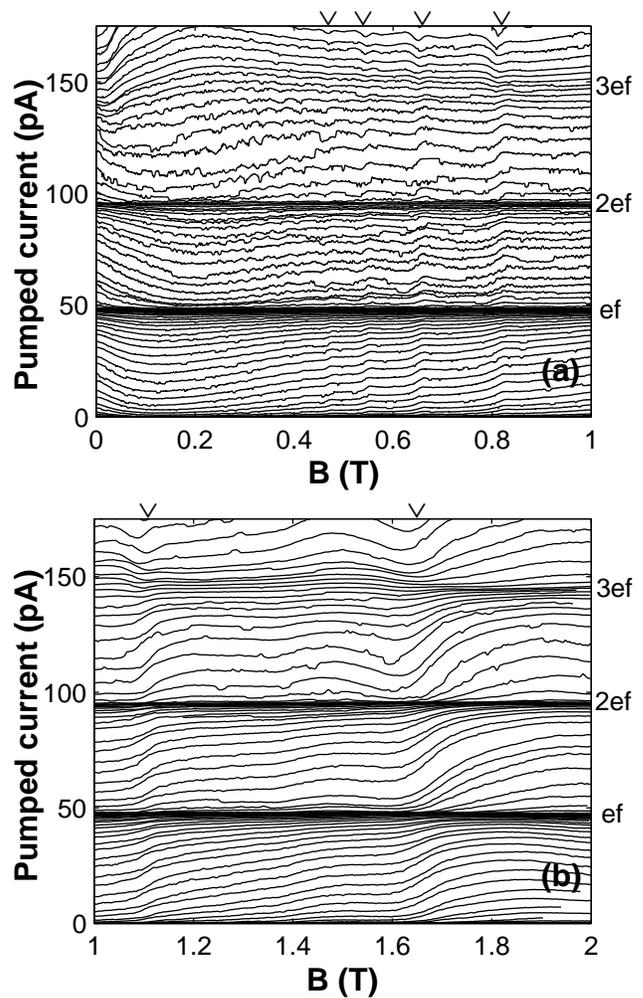}
\caption{\label{fig:mag1} Pumped current as a function of applied perpendicular $B$ field. Each line represents a different voltage $V_{2}$, explained in more detail in the text. The v's along the upper borders highlight features of interest. A sweep rate of 5$\,$T/hr and 20$\,$T/hr was used in a) and b) respectively.}
\end{figure}

Figure~\ref{fig:plat} shows pumped current with no applied magnetic field. The first two plateaus are clearly visible and correspond to the pumping of one and two electrons through the dot respectively. Higher $n$ plateaus are generally less well defined\cite{Blumenthal2007}. Dashed lines represent the expected current for the first three plateaus. When the $B$-field is applied, changes in the pumped current as a function of magnetic field are seen. Figure~\ref{fig:mag1} shows the magnetic field dependence for pumped current at different values of $V_{2}$, incremented in steps of $670\mu$V. $V_{1}$ was fixed at $-0.140\,$V. Plateaus corresponding to a quantized number $n$ of electrons pumped per cycle manifest as darkened areas in the plot where several curves condense. Switching events, commonly referred to as random telegraph signals, are noticeable for the voltages corresponding to transitions between plateaus. For these voltages, the total pumped current is more sensitive to changes in the background potential. These events are less clear in Fig.~\ref{fig:mag1} (b), where the $B$ field sweep rate was increased from $5\,$T/hr to $20\,$T/hr.

In Fig.~\ref{fig:mag1} (a) and (b) it can be seen that the plateaus for $ef$, $2ef$ and $3ef$ are visible for magnetic fields in the range of $0\,$T to $2\,$T. This is significantly different to observations made by Cunningham \emph{et al}\cite{Cunningham2000}. For SAW devices the $ef$ plateau was no longer visible for magnetic fields above $250\,$mT and higher plateaus ($2ef$ and $3ef$) disappeared at even lower magnetic fields.

In Fig.~\ref{fig:mag1} (a) and (b) the pumped current for a fixed voltage $V_{2}$ gradually increases as a function of magnetic field. For voltages $V_2$ such that the pumped current is not quantized, increasing the magnetic field causes the current to rise towards the quantized value. The plateaus are therefore becoming longer in $V_{2}$, indicating an improvement in the robustness of the quantization as the applied $B$ field is increased.

This effect is particularly apparent in the field range of 1.7-2$\,$T, where the $3ef$ plateau becomes considerably more pronounced. This may be due to the stronger confinement of the captured electrons in the dot as the magnetic field is increased. The eigenstate of the many-particle system is classified by one angular momentum quantum number ($J$) which is the sum of the single-electron quantum numbers\cite{maksym}.  Because of electron-electron interaction, the ground state energy increases with this total angular momentum quantum number. The extent of the ground state wave function in the radial direction, i.e., the radius of the orbit, is determined by the magnitude of $a$ where $a^2=(\hbar/m^\ast)(\omega_c^2+4\omega_0^2)^{-1/2}$. In this notation, $m^\ast$ is the electron effective mass, $\omega_c$ is the cyclotron frequency and $\omega_0$ is the angular frequency for the harmonic potential of the quantum dot. As the magnetic field increases, $a$ decreases, thus making the total wave function more confined close to the center of the dot and less likely for an electron to escape. As a result, the probability for electrons to tunnel back into the source after they have been captured (as discussed in Ref\cite{Blumenthal2007}) is reduced, and therefore quantization in the pumped current is improved.

Figure~\ref{fig:mag1} (a) and (b) also show features at certain values of magnetic field, periodic in $1/B$, as indicated by the v's along the upper borders. We remark that the positions of these features correspond to the cyclotron radius $R_c=m^* v_{F}/eB$ being an integer multiple of 21~nm (where the Fermi velocity $v_F$ is assumed to be that of the bulk). This length scale is comparable to the average size of the QD, as estimated from a time-dependent potential calculated using the lithographic device design\cite{davies}. Similar features in the current pumped by a SAW were attributed to commensurability oscillations in the 2DEG adjacent to the SAW channel\cite{Cunningham2000,gg}. A similar explanation for the features in our data is unlikely as there is no 2D region near to the electron pump. We do not have a complete explanation for the origin of these features at present.

\begin{figure}[t]
\includegraphics[width=0.45\textwidth]{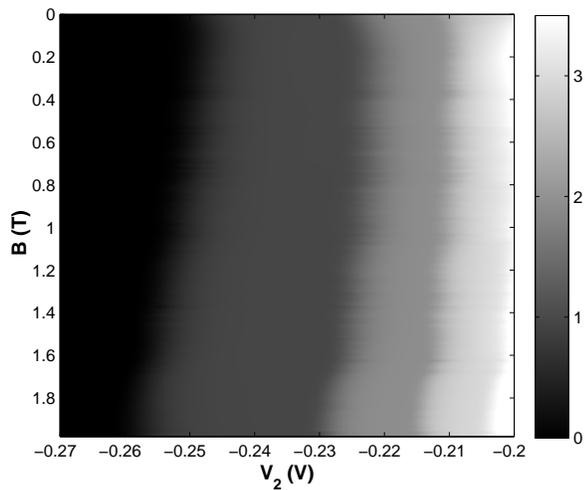}
\caption{\label{fig:mag2} Evolution of the quantized pumped current plateaus in applied $B$ field. The numbers in the gray scale bar indicate the number of electrons pumped per cycle.}
\end{figure}

\begin{figure*}[!b]
\includegraphics[width=1\textwidth]{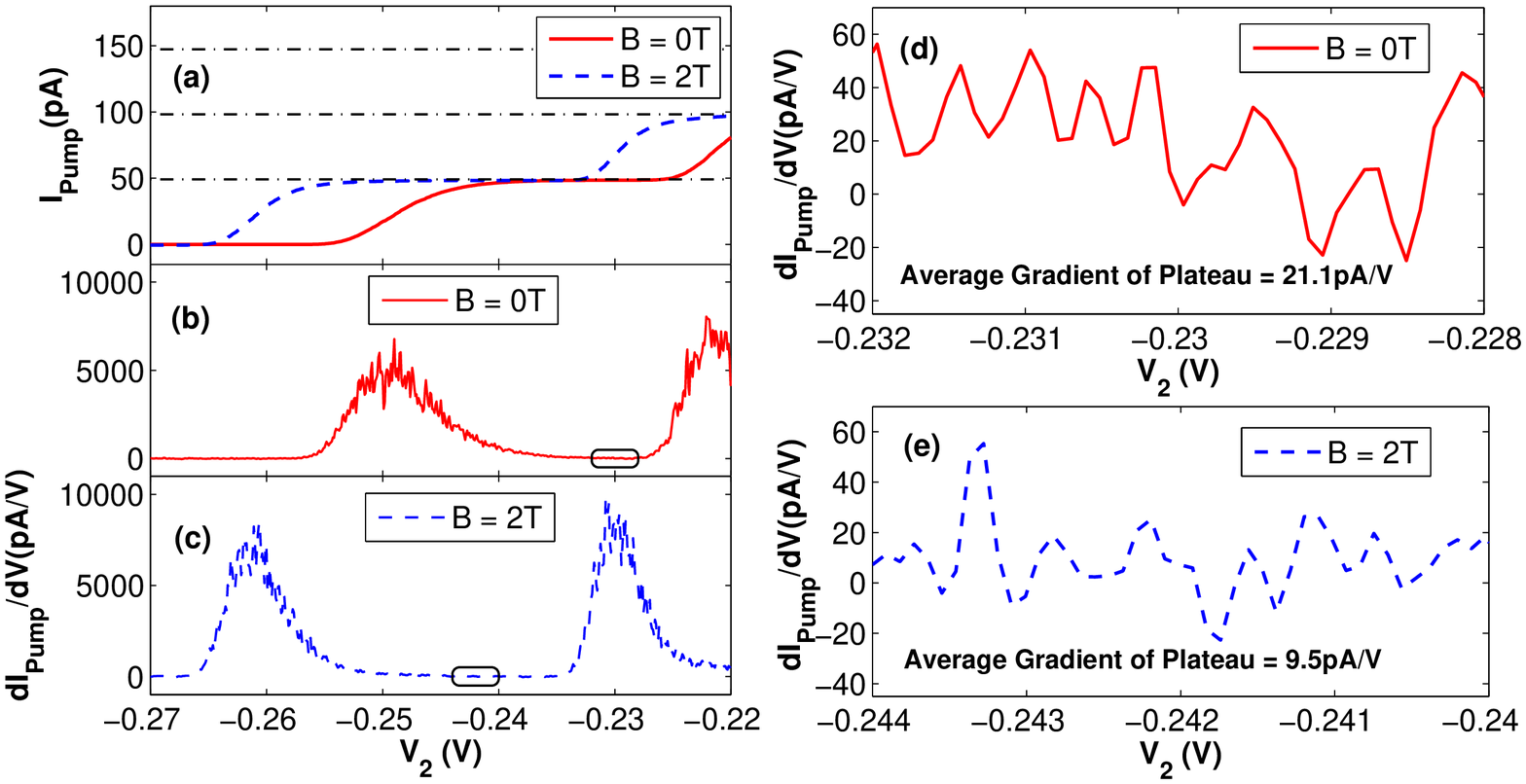}
\caption{\label{fig:mag3} (Color online) Pumped current for the first plateau. a) shows the pumped current at 0T (red solid curve) and 2.5T (blue dashed curve). Dot dashed lines correspond to the expected plateau values. b) and c) show the numerical derivative of the pumped current. d) and e) are expanded regions indicated by the black boxes in in b) and c) over which the gradient of each plateau were calculated.}
\end{figure*}

Figure~\ref{fig:mag2} shows the evolution of the data in Fig.~\ref{fig:plat} under an applied $B$ field. It is clear from this Figure that the plateaus become longer, again showing the increased robustness of the pumped current due to the magnetic field.

Figures~\ref{fig:mag3} (a) through~\ref{fig:mag3} (e) contain a set of plots with a common gate voltage axis that will be used to present a qualitative analysis of the accuracy of the pumped current. Figure~\ref{fig:mag3} a) shows the first plateau of pumped current in zero field (red solid curve) and in a field of 2.5T (blue dashed curve). The numerical derivative for each plot in Fig.~\ref{fig:mag3} a) was taken, with the results plotted in Fig.~\ref{fig:mag3} b) and c) for zero field and 2.5T respectively. On first observation the increase in the length of the plateau in a field of 2.5T can be seen, as was previously discussed. Regions of minimum gradient are shown by the black boxes in Fig.~\ref{fig:mag3} b) and c). The size of box was chosen to encompass a region where the numerical derivative is at a minimum in zero field. A magnified plot of these regions is presented in Fig.~\ref{fig:mag3} d) and e). The range of gate voltages were chosen to be the same in order to draw an accurate comparison in the gradient of the plateau. From Fig.~\ref{fig:mag3} we determine an average gradient for the first plateau of 21.1$\,$pA/V in zero field which improves to a gradient of 9.5$\,$pA/V for a field of 2.5 T. Janssen and Hartland\cite{janssen} have demonstrated an empirical relation between the slope of the plateau and the accuracy of quantization. The present results show an improvement of $\sim$55\% in the current quantization with the application of a perpendicular magnetic field of 2.5T

In conclusion, we have presented experimental results of the effects observed when a high-frequency electron pump is exposed to a perpendicular magnetic field. Strong improvements in the plateaus of quantized current were clearly seen. The addition of a perpendicular field introduces an additional confinement to the electrons captured by the dynamic dot. Such confinement has had a positive effect, leading to increased robustness and improved quantization. Further work is required to fully appreciate the physical effects a magnetic field has on these types of devices. Such findings would have a positive influence on the many applications proposed for these robust and dynamic electron pumps.\\

We thank Dr Chris Ford, Dr Masaya Kataoka, Jonathan Griffiths, Lee Bassett, Dr Simon Chorley, Jonathan Prance and Karl Petersson for useful discussions. SJW acknowledges support from the EPSRC and Toshiba Research Europe Ltd. The work of MDB was supported by the UK National Measurement System's Quantum Metrology Programme. The work of GG was supported by contract FA9453-07-C-0207 of AFRL. CN acknowledges support from the EPSRC QIP IRC (GR/S82176/01).


\begin{thebibliography}{99}

\footnotesize{

\bibitem{geerligs1990} L.~J.~Geerligs, V.~F.~Anderegg, P.~A.~M.~Holweg, J.~E.~Mooij, H.~Pothier, D.~Esteve, C.~Urbina, and M.~H.~Devoret, Phys. Rev. Let. {\bf 64}, 2691 (1990).

\bibitem{andregg1990} V.~F.~Anderegg, L.~J.~Geerligs, J.~E.~Mooij, H.~Pothier, D.~Esteve, C.~Urbina, and M.~H.~Devoret, Physica B {\bf 61}, 165-166 (1990).

\bibitem{kouwenhoven1PBI} L.~P.~Kouwenhoven, A.~T.~Johnson, N.~C.~van der Vaart, C.~J.~P.~M.~Harmans and C.~T.~Foxon Phys. Rev. Lett. {\bf 67}, 1626 (1991).

\bibitem{pothier1PBI} H.~Pothier, P.~Lafarge, C.~Urbina, D.~Esteve, and M.~H.~Devoret, Europhys. Lett. {\bf 17}, 249 (1992).

\bibitem{keller} M.~W.~Keller, J.~M.~Martinis, N.~M.~Zimmerman, and A.~H.~Steinbach, Appl. Phys. Lett. {\bf 69}, 1804 (1996).

\bibitem{Shilton1996} J.~M.~Shilton, V.~I.~Talyanskii, M.~Pepper, D.~A.~Ritchie, J.~E.~F.~Frost, C.~J.~B.~Ford, C.~G.~Smith, and G.~A.~C.~Jones, J. Phys. Cond Mat {\bf 8}, L531 (1996).

\bibitem{Thornton1986} T.~J.~Thornton, M.~Pepper, H.~Ahmed, D.~Andrews, and G.~J.~Davies, Phys. Rev. Lett. {\bf 56}, 1198 (1986).

\bibitem{Blumenthal2007} M.~D.~Blumenthal, B.~Kaestner, L.~Li, S.~Giblin, T.~J.~B.~M.~Janssen, M.~Pepper, D.~Anderson, G.~Jones, and D.~A.~Ritchie, Nature Physics {\bf 3}, 343 (2007).

\bibitem{kaestner:153301} B.~Kaestner, V.~Kashcheyevs, S.~Amakawa, M.~D.~Blumenthal, L.~Li, T.~J.~B.~M.~Janssen, G.~Hein, K.~Pierz, T.~Weimann, U.~Siegner and H.~W.~Schumacher, Phys. Rev. B {\bf 77}, 153301 (2008).

\bibitem{kaestner:192106} B.~Kaestner, V.~Kashcheyevs, G.~Hein, K.~Pierz, U.~Siegner, and H.~W.~Schumacher, Appl. Phys. Lett. {\bf 92}, 192106 (2008).

\bibitem{Cunningham2000} J.~Cunningham, V.~I.~Talyanskii, J.~M.~Shilton, M.~Pepper, A.~Kristensen, and P.~E.~Lindelof, Physical Review B {\bf 62}, 1564 (2000).

\bibitem{maksym} P.~A.~Maksym and Tapash Chakraborty, Phys. Rev. Lett. {\bf 65}, 108  (1990).

\bibitem{davies} J.~H.~Davies, I.~A.~Larkin and E.~V.~Sukhorukov, J. Appl. Phys. {\bf 77}, 4505 (1995).

\bibitem{gg} Natalya Zimbovskaya and Godfrey Gumbs, J. Phys. Cond. Matt. {\bf 13}, L1-L8 (2001).

\bibitem{janssen} T.~J.~B.~M.~Janssen and A.~Hartland, IEEE Proc.-Sci. Meas. Technol. {\bf 147}, 174 (2000).

}


\end{thebibliography}
\end{document}